\documentclass[preprint,showpacs,nofootinbib,preprintnumbers,amsmath,amssymb]{revtex4}

\usepackage{graphicx}
\usepackage{mathrsfs}
\usepackage{amssymb}
\usepackage{bm}

\begin{document}

\title{Collins effect in semi-inclusive deep inelastic scattering  process with a $^3$He target}

\newcommand*{\PKU}{School of Physics and State Key Laboratory of Nuclear Physics and
Technology, \\Peking University, Beijing 100871,
China}\affiliation{\PKU}
\newcommand*{\CHEP}{Center for High Energy
Physics, Peking University, Beijing 100871,
China}\affiliation{\CHEP}

\author{Jun She}\affiliation{\PKU}
\author{Bo-Qiang Ma}\email{mabq@pku.edu.cn}\affiliation{\PKU}\affiliation{\CHEP}

\begin{abstract}
We re-examine our previous calculation on the Collins effect in
semi-inclusive deeply inelastic scattering (SIDIS) process with a
$^3$He target, and find that our previous treatment on the dilution
factors may cause the results larger than the realistic situation.
We thus modify our calculation in an improved treatment with an
updated prediction on the $\sin(\phi_h+\phi_S)$ asymmetry for the
JLab 12 GeV under the transverse momentum dependent (TMD)
factorization framework. Meanwhile, we also provide the prediction
of such asymmetry for the JLab 6 GeV and the prediction of the
$\sin(3\phi_h-\phi_S)$ asymmetry related to pretzelosity.
\end{abstract}

 \maketitle

\section{introduction}
The transverse spin structure of the nucleon is a hot issue in spin
physics. Single spin asymmetry (SSA) through semi-inclusive deeply
inelastic scattering (SIDIS) process with a transversely polarized
target provides us with a plausible way to explore the internal
structure of the nucleon. In such process a number of form factors
with different azimuthal angle dependence are measurable. These form
factors can be interpreted as the convolution of the parton
distribution functions and fragmentation functions. According to the
transverse momentum dependent (TMD) factorization~\cite{tmd}, which
is intuitively a natural generalization of the collinear case by
taking into account the transverse momentum, eight TMD distribution
functions are needed to describe the spin structure of the nucleon
at the leading twist. Among the eight, some of them relating to the
transverse spin effect have attracted many interests in recent
years, such as the Sivers, transversity and newly pretzelosity
functions, which can manifest themselves through the
$\sin(\phi_h-\phi_S)$, $\sin(\phi_h+\phi_S)$ and
$\sin(3\phi_h-\phi_S)$ asymmetries, respectively in the SIDIS
process.

The HERMES Collaboration has already published their data on the
proton target~\cite{hermes} with non-vanishing $\sin(\phi_h-\phi_S)$
and $\sin(\phi_h+\phi_S)$ asymmetries. Some unexpected phenomena
were also observed, for example, the $\sin(\phi_h+\phi_S)$ asymmetry
for $\pi^-$ production is larger than that for $\pi^+$ production. A
possible explantation for this is that the corresponding unfavored
fragmentation function is comparable with or even larger than the
favored one~\cite{Vogelsang2005}, i.e.,
$|H_{1\mathrm{unf}}^\perp|\gtrsim |H_{1\mathrm{fav}}^\perp|$, where
$H_1^\perp$ is the so called Collins function~\cite{Collins}. This
relation is now widely adopted in parametrizations, but is still a
puzzling theoretical problem. The COMPASS
Collaboration~\cite{compass} has also made the measurement with the
deuteron target, which can be viewed as a combination of a proton
and a neutron, and observed nearly zero asymmetries for both
$\sin(\phi_h-\phi_S)$ and $\sin(\phi_h+\phi_S)$ asymmetries. This
could be accounted for as the cancelation between the $u$ and $d$
quarks. So a direct measurement with the neutron target is ideal to
address this issue. However, there is no free neutron target for the
experiment, so usually the $^3$He target is used as a substitution.
$^3$He is a spin half nucleus, composed by two protons and one
neutron, with the polarizations of the two protons anti-paralleled,
thus $^3$He can be viewed as a nearly free neutron. Jefferson Lab
(JLab) is performing the measurement with a $^3$He target under 6
GeV, and will upgrade their plan by a 12 GeV beam energy in the near
future~\cite{Gao2010}. Such plan can be considered as performing a
measurement for the purpose to focus attention on the neutron
target.

However, after taking into account other effects, $^3$He cannot be
viewed as a pure neutron strictly. How to connect the experiment
result of the $^3$He target with the free neutron case needs to be
carefully considered. Some theoretical work~\cite{Scopetta} already
investigated the problem, and the following relation is often used,
\begin{eqnarray}
A_{^3\mathrm{He}}=p_n\cdot f_n\cdot A_n+2p_p\cdot f_p\cdot
A_p,\label{A1_He}
\end{eqnarray}
where $p_{p(n)}$ are the effective nucleon polarizations, with
$p_p\approx-0.028$ and $p_n\approx0.86$ as suggested in
Ref.~\cite{Scopetta}. $f_{p(n)}$ are the dilution factors with the
definitions
\begin{eqnarray}
f_{p(n)}(x,z,\bm{P}_{h\perp})=\frac{\int\sum_q e_q^2
f_1^{q,p(n)}(x,\bm{p}_\perp)
D_1^{q,h}(z,\bm{k}_\perp)}{\sum_{N=p,n}\int\sum_q e_q^2
f_1^{q,N}(x,\bm{p}_\perp) D_1^{q,h}(z,\bm{k}_\perp)}.
\end{eqnarray}
We have simply generalized the expression above into the TMD case,
and the integrals in the numerator and denominator are performed
over all the phase space independently. The validity of
Eq.~(\ref{A1_He}) was discussed in Refs.~\cite{Scopetta}. In our
previous paper~\cite{Huang2007}, we already used this formula to
predict the $^3$He asymmetry, but with a constant value estimation
for the dilution factors, adopted $f_p\approx0.34$ and
$f_n\approx0.32$ around $z\approx0.5$ as proposed~\cite{pac23}. Now
we have re-examined our calculation, and find that this rough
estimation is not adequate to give the proper prediction. In this
brief note, we deal with the problem in an improved treatment with
the dilution factors varying with kinematic variables, meanwhile we
update our prediction from the following aspects:
\begin{itemize}
\item We present predictions not only for JLab 12 GeV, but also for JLab
6 GeV to confront with the data that will come soon.
\item We calculate under the TMD framework, and show the
asymmetries depending on $x$, $z$ and $P_{h\perp}$.
\item We multiply only the azimuthal angle dependence as the
weighting function to extract the corresponding asymmetry, rather
than multiply an extra momentum dependent factor as we did in our
previous paper. The former treatment is often used in many
experiments, while the latter one is not so widely adopted, although
it can avoid the tangle of the quark internal momenta from a
theoretical viewpoint.
\item We also give the asymmetry related to pretzelosity,
another chiral-odd but T-even TMD distribution, which also couples
with the Collins functions in the SIDIS process.
\end{itemize}

\section{Numerical calculation}
We write down the cross-section with the Collins effect
terms~\cite{Kotzinian1995,Bacchetta2007}
\begin{eqnarray}
\frac{d^6\sigma_{UT}}{dxdyd\phi_Sdzd^2\bm{P}_{h\perp}}=&&\frac{2\alpha^2}{sxy^2}\{
(1-y+\frac{1}{2}y^2)F_{UU}+S_\perp(1-y)\nonumber\\
&&\times[\sin(\phi_h+\phi_S)F^{\sin(\phi_h+\phi_S)}_{UT}
+\sin(3\phi_h-\phi_S)F^{\sin(3\phi_h-\phi_S)}_{UT}]+...\},
\end{eqnarray}
with
\begin{eqnarray}
&&F_{UU}=\mathcal{F}[f_1 D_1],\\
&&F^{\sin(\phi_h+\phi_S)}_{UT}=\mathcal{F}[-\frac{\hat{\bm{h}}\cdot\bm{k}_\perp}{M_h}
h_{1} H_1^\perp],\\
&&F^{\sin(3\phi_h-\phi_S)}_{UT}=\mathcal{F}[\frac{2(\hat{\bm{h}}\cdot\bm{p}_\perp)
(\bm{p}_\perp\cdot\bm{k}_\perp)
+\bm{p}_\perp^2(\hat{\bm{h}}\cdot\bm{k}_\perp)-
4(\hat{\bm{h}}\cdot\bm{p}_\perp)^2(\hat{\bm{h}}\cdot\bm{k}_\perp)}{2M_N^2M_h}
h_{1T}^\perp H_1^\perp],~~~~~
\end{eqnarray}
where a compact notation
\begin{eqnarray}
\mathcal{F}[\omega f D]=\sum_q e_q^2\int d^2\bm{p}_\perp
d^2\bm{k}_\perp
\delta^2(\bm{p}_\perp-\bm{k}_\perp-\bm{P}_{h\perp}/z)\nonumber\\
\omega(\bm{p}_\perp,\bm{k}_\perp)f^q(x,\bm{p}_\perp^2)D^q(z,z^2\bm{k}_\perp^2)
\end{eqnarray}
is used. We might as well take the $\sin(\phi_h+\phi_S)$ asymmetry
as an example, and the other asymmetries can be analyzed in the same
way. The asymmetry for a $^3$He target can be calculated from
\begin{eqnarray}
A_{^3\mathrm{He}}^{\sin(\phi_h+\phi_S)}(x,z,\bm{P}_{h\perp})=\frac{\int
dy \frac{2\alpha^2}{sxy^2}(1-y)
\mathcal{F}[-\frac{\hat{\bm{h}}\cdot\bm{k}_\perp}{M_h}
h_{1}^{^3\mathrm{He}} H_1^\perp]} {\int dy
\frac{2\alpha^2}{sxy^2}(1-y+\frac{1}{2}y^2)\mathcal{F}[f_1^{^3\mathrm{He}}
D_1]},\label{A2_He}
\end{eqnarray}
while for a proton or a neutron target, this can be written as
\begin{eqnarray}
A_{p(n)}^{\sin(\phi_h+\phi_S)}(x,z,\bm{P}_{h\perp})=\frac{\int dy
\frac{2\alpha^2}{sxy^2}
(1-y)\mathcal{F}[-\frac{\hat{\bm{h}}\cdot\bm{k}_\perp}{M_h}
h_{1}^{p(n)} H_1^\perp]} {\int dy
\frac{2\alpha^2}{sxy^2}(1-y+\frac{1}{2}y^2)\mathcal{F}[f_1^{p(n)}
D_1]}.\label{A_N}
\end{eqnarray}
Now we rewrite the expression for the dilution factors in a more
complete form
\begin{eqnarray}
f_{p(n)}(x,z,\bm{P}_{h\perp})=\frac{\int dy
\frac{2\alpha^2}{sxy^2}(1-y+\frac{1}{2}y^2)\mathcal{F}[f_1^{p(n)}
D_1]} {\int dy
\frac{2\alpha^2}{sxy^2}(1-y+\frac{1}{2}y^2)\mathcal{F}[(2f_1^p+f_1^n)
D_1]}.\label{dilution}
\end{eqnarray}
Substituting Eqs.~(\ref{A_N}) and (\ref{dilution}) into
Eq.~(\ref{A1_He}), we have
\begin{eqnarray}
A_{^3\mathrm{He}}^{\sin(\phi_h+\phi_S)}(x,z,\bm{P}_{h\perp})=\frac{\int
dy \frac{2\alpha^2}{sxy^2}
(1-y)\mathcal{F}[-\frac{\hat{\bm{h}}\cdot\bm{k}_\perp}{M_h} (2p_p
h_{1}^p+p_n h_1^n) H_1^\perp]} {\int dy
\frac{2\alpha^2}{sxy^2}(1-y+\frac{1}{2}y^2)\mathcal{F}[(2f_1^p+f_1^n)
D_1]}.\label{A3_He}
\end{eqnarray}
Comparing Eq.~(\ref{A3_He}) with Eq.~(\ref{A2_He}), we find that it
implies the equalities
\begin{eqnarray}
&&h_1^{^3\mathrm{He}}=2p_p h_1^p +p_n h_1^n,\nonumber\\
&&f_1^{^3\mathrm{He}}=\sum_{N=p,n} f_1^N=2f_1^p+f_1^n,
\label{pdf_He}
\end{eqnarray}
which can be understood as the impulse approximation to treat
scattering on both unpolarized and polarized $^3$He as an incoherent
sum of scattering on free nucleons. The first line of the above
equalities shows that the transversity distribution of $^3$He is a
linear combination of the free nucleon transversity with the
effective polarizations as their coefficients. We take it as a
reasonable assumption that can be generalized to other polarized
distributions such as pretzelosity. The second line of the
equalities implies that for the unpolarized distribution, $^3$He can
be treated as a sum of free nucleons. Strictly speaking, this is not
accurate because the nuclear effects such as the EMC
effect~\cite{emc} are neglected. However, given that $^3$He is a few
body system, the EMC effect is not significant, and thus we could
admit that the second line of the above equalities holds as an
approximation. So Eq.~(\ref{A1_He}) is equivalent to
Eq.~(\ref{pdf_He}). We could also understand it in the following
way: we view Eq.~(\ref{pdf_He}) as our general assumption, then
starting from this equation, we can derive Eq.~(\ref{A1_He}) from
the definition of the asymmetries for the proton (neutron) and
$^3$He. However, we should not take the dilution factors as fixed
constants as in our previous calculation.

In the next calculation, the TMD distributions will be derived from
an SU(6) quark-diquark model as we used in our previous
papers~\cite{Ma,She2009}, with the Collins function and the
unpolarized fragmentation function parametrized from Anselmino {\it
et al.}~\cite{Anselmino2009} and Kretzer {\it et
al.}~\cite{Kretzer2001}, respectively. The kinematics for the 6 and
12 GeV are in Table \ref{kin}.
\begin{table}
\begin{center}
\caption{kinematics} \label{kin}
\begin{tabular}{|c|c|}
\hline\hline
~~~~~~6~GeV~~~~~~&~~~~~~12 GeV~~~~~~ \\
\hline\hline ~~~~~~$1.3$GeV$^2<Q^2<3.1$GeV$^2$~~~~~~&~~~~~$Q^2>1$GeV$^2$~~~~~~\\
\hline ~~~~~~$5.4$GeV$^2<W^2<9.3$GeV$^2$~~~~~~&~~~~~$W^2>2.3$GeV$^2$~~~~~~\\
\hline ~~~~~$0.13<x<0.4$~~~~~~&~~~~~~$0.05<x<0.65$~~~~~\\
\hline ~~~~~$0.68<y<0.86$~~~~~~&~~~~~~$0.34<y<0.9$~~~~~\\
\hline ~~~~~$0.46<z<0.59$~~~~~~&~~~~~~~$0.3<z<0.7~~~~~~$ \\
\hline
\end{tabular}
\end{center}
\end{table}

\begin{figure}
\includegraphics[scale=0.8]{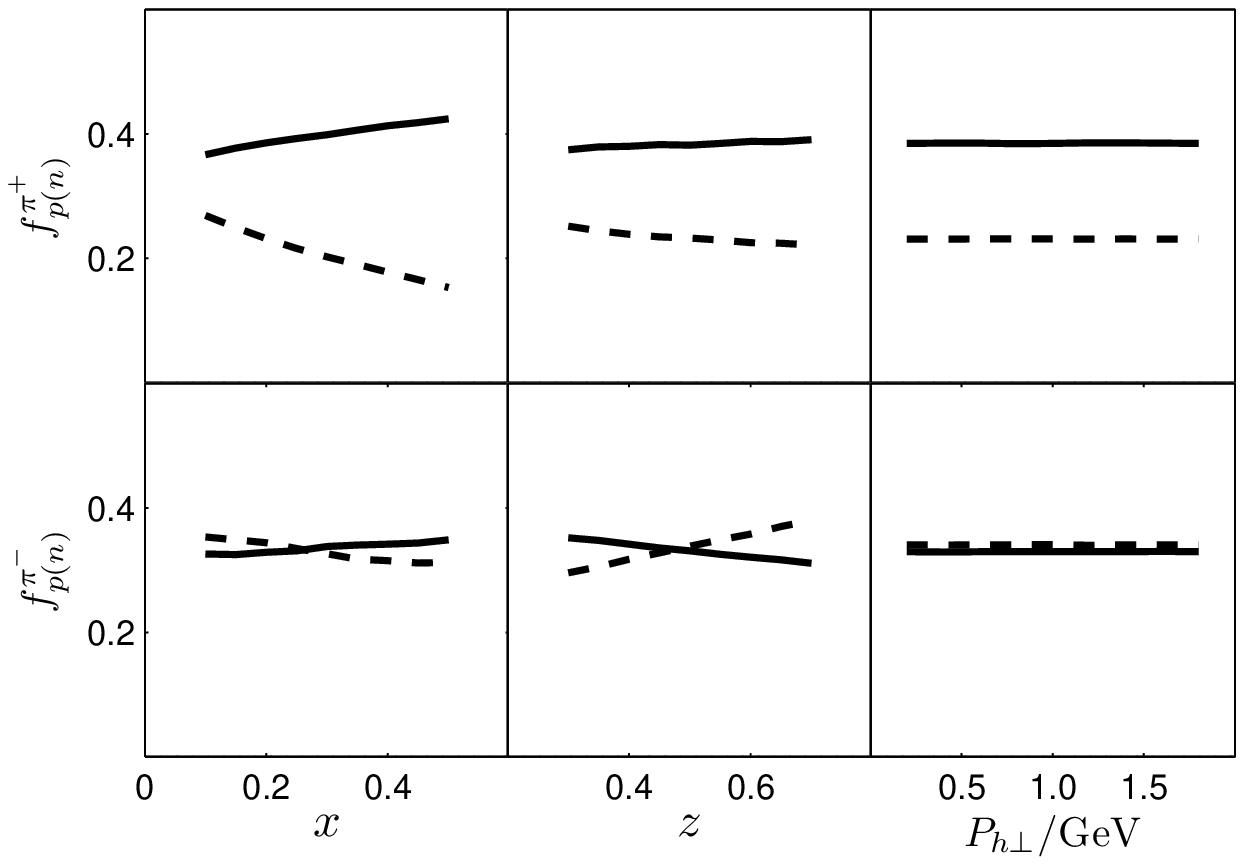}
\caption{Dilution factors for $\pi$ production off a $^3$He target
at JLab 6 GeV kinematics. The solid and dashed curves are the
results for the proton and neutron dilution effects, respectively.}
\label{dilution6}
\includegraphics[scale=0.8]{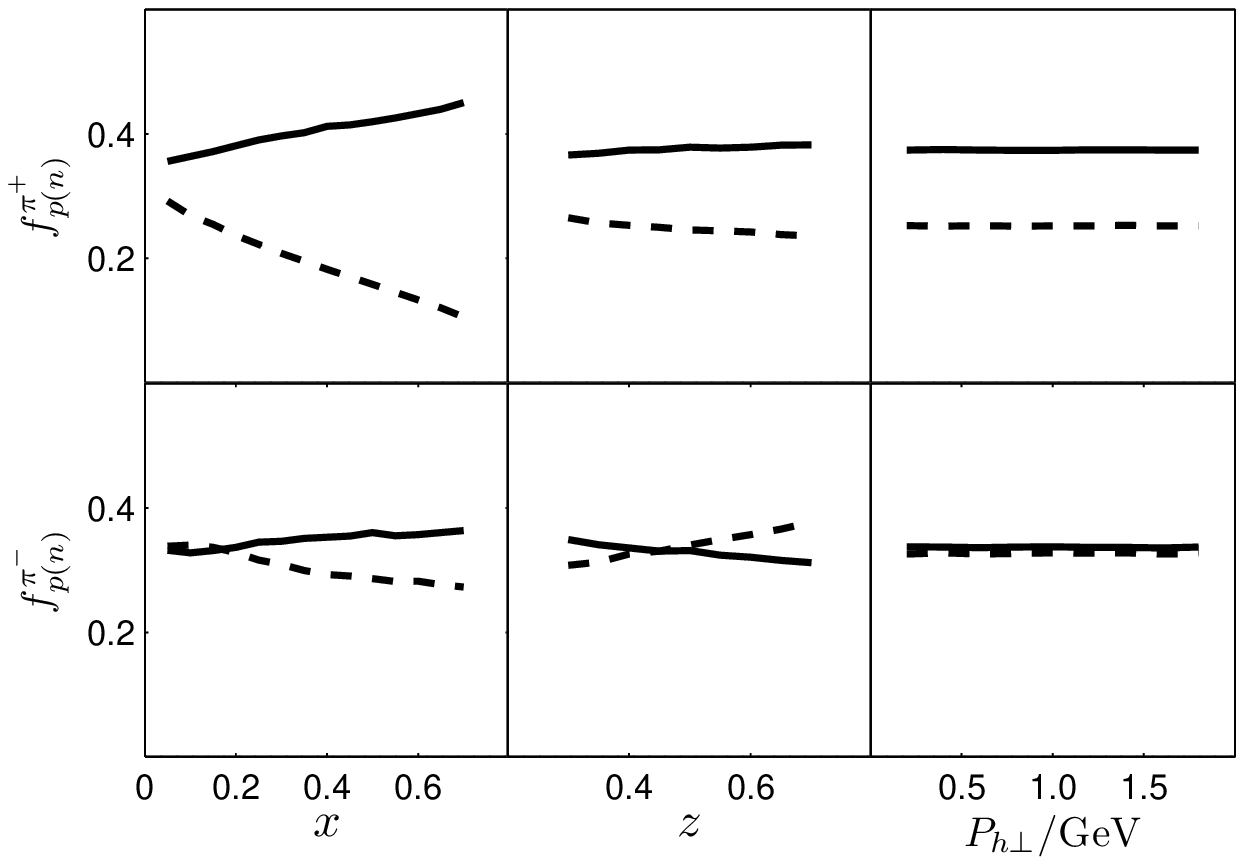}
\caption{Similar as Fig.~\ref{dilution6}, but for the 12 GeV
kinematics.} \label{dilution12}
\end{figure}

First, we will present our numerical results on the dilution factors
under 6 and 12 GeV kinematics in Fig.~\ref{dilution6} and
Fig.~\ref{dilution12}, respectively.
From the two figures, we find that the dilution factors vary very
little with $z$ and $P_{h\perp}$, but vary significantly with $x$,
especially for the $\pi^+$ production. In our previous
calculation~\cite{Huang2007}, we used the estimation
$f_p\approx0.34$ and $f_n\approx0.32$ as proposed~\cite{pac23}. We
are mainly interested in the $x$ dependence, and could make a
comparison with our results. For the $\pi^-$ production, our results
are very close with that in the proposal~\cite{pac23}. However, for
the $\pi^+$ production, we give a much smaller dilution factor for
neutron and a larger one for proton, especially when $x$ increases.
This can be easily understood that for the $\pi^+$ production,
$u\rightarrow\pi^+$ is favored and $d\rightarrow\pi^+$ is unfavored.
Since the $^3\mathrm{He}$ target can be viewed as a nearly free
neutron target, we expect that our updated results of the
asymmetries for the $\pi^+$ production might be suppressed due to
this new treatment with the dilution factors.

\begin{figure}
\includegraphics[scale=0.8]{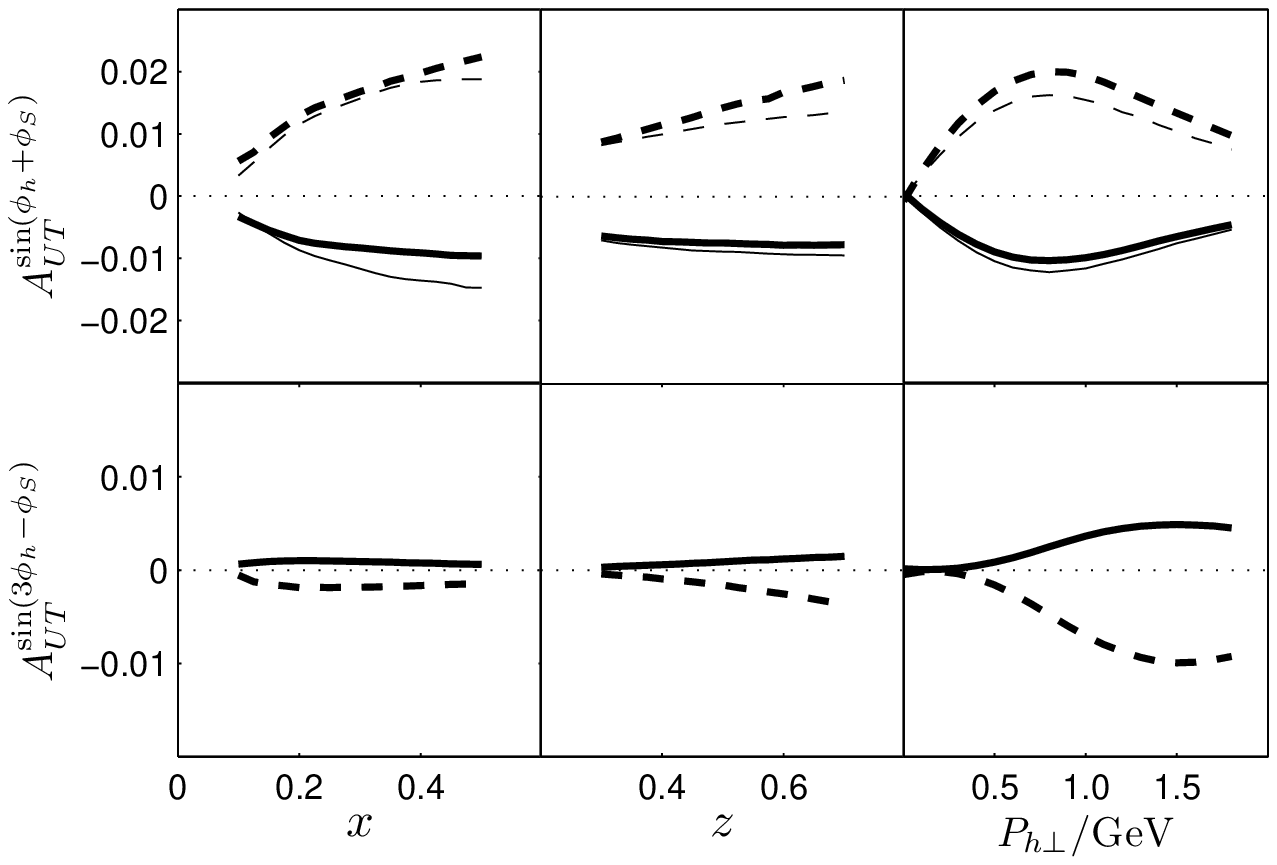}
\caption{$\sin(\phi_h+\phi_S)$ and $\sin(3\phi_h-\phi_S)$
asymmetries for $\pi$ production off a $^3$He target at JLab 6 GeV
kinematics. The solid and dashed curves are the results for the
$\pi^+$ and $\pi^-$ production, respectively. The thin curves are
the results in which we make the assumption that the dilution
factors are constants as we did in our previous paper.} \label{6GeV}
\includegraphics[scale=0.8]{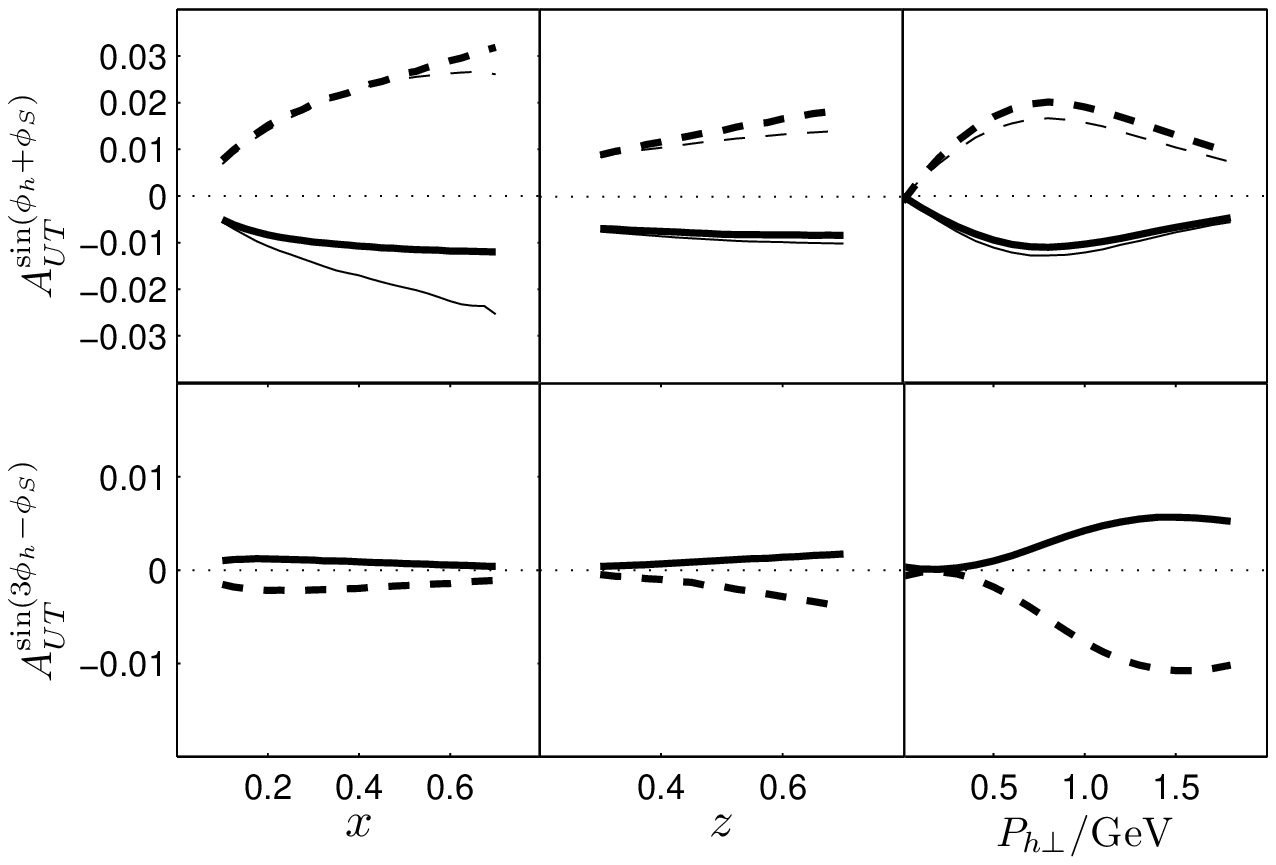}
\caption{Similar as Fig.~\ref{6GeV}, but for the 12 GeV kinematics.}
\label{12GeV}
\end{figure}

Fig.~\ref{6GeV} and Fig.~\ref{12GeV} shows the $\sin(\phi_h+\phi_S)$
and the $\sin(3\phi_h-\phi_S)$ asymmetries at JLab 6 GeV and 12 GeV,
respectively.
From the figures, we can see that the asymmetries for the $^3$He
target are rather small. This can be accounted for by the
suppression of the dilution factors. We deal with the dilution
factors in two approaches, one is to use the constant value
estimation, i.e. $f_n\approx0.32$ and $f_p\approx0.34$ as was
supposed to hold around $z\approx0.5$~\cite{pac23} and used in our
previous calculation~\cite{Huang2007}, and the other is to calculate
from the definition in Eq.~(\ref{dilution}), where the dilution
factors vary with $x$,  $z$ and $P_{h\perp}$. We find that the
results of the two approaches are very close to each other for the
$z$ and $P_{h\perp}$ dependence. However, it has some difference for
the $x$ dependence. As we expected, the asymmetry for the $\pi^-$
production in both approaches are also very close to each other. But
for the $\pi^+$ production, the asymmetry in the latter approach is
indeed smaller than that in the former approach according to our
analysis in the above. Such small results might be achieved in JLab
with a high precision measurement, so we should be careful about
this effect caused by the dilution factors. Also we plot the
$\sin(3\phi_h-\phi_S)$ asymmetry in the lower panels in
Fig.~\ref{6GeV} and Fig.~\ref{12GeV}. This asymmetry is indeed much
smaller than the $\sin(\phi_h+\phi_S)$ asymmetry as we expected in
our previous calculation~\cite{She2009}. This implies that it might
be more difficult to measure the $\sin(3\phi_h-\phi_S)$ asymmetry
related to the pretzelosity, unless a high precision can be
achieved. It is suggested~\cite{She2009} that the asymmetry can be
amplified by introducing a transverse momentum cut in data analysis.

\section{Conclusion}
Eq.~(\ref{A1_He}) was supposed as a useful formula to extract the
neutron information in the experiment. Using mean values for the
dilution factors in the calculation, we find that such treatment
might be approximately valid for the $z$ and $P_{h\perp}$
dependence, and even for the $x$ dependence in the $\pi^-$
production, but cannot be applied for the $x$ dependence in the
$\pi^+$ production. In this note, we performed the calculation
starting from the definition (\ref{dilution}), which is demonstrated
to be equivalent to the relation~(\ref{pdf_He}), and find that this
approach may cause smaller results of asymmetries than our previous
calculation~\cite{Huang2007} for the $\pi^+$ production process. So
we suggest that in extracting the neutron information, it is not
adequate to use the mean values for the dilution factors. Unreliable
treatment of dilution factors may bring unreliable information of
the neutron quantities extracted from  $^3$He data. Therefore it is
important to find reliable treatment of dilution factors in
extracting the neutron asymmetries from  $^3$He experiments. Due to
the suppression from the dilution factors, all the $^3$He
asymmetries are predicted to be very small, so high precision
measurements are demanded.

\section*{Acknowledgement}
This work is supported by National Natural Science Foundation of
China (Nos.~10905059, 11021092, 10975003, and 11035003).

\end{document}